\documentstyle[twocolumn,pre,aps,epsf]{revtex}
\newcommand{\be}{\begin{equation}}
\newcommand{\ee}{\end{equation}}
\newcommand{\bear}{\begin{eqnarray}}
\newcommand{\eear}{\end{eqnarray}}

\def\Tc{{T_{\rm c}}}
\def\al{\alpha}
\def\bt{\beta}
\begin{document}
\twocolumn[\hsize\textwidth\columnwidth\hsize\csname @twocolumnfalse\endcsname

\title{Unconventional odd-parity superconductivity in the ladder compounds}
\author{Mahito Kohmoto and Masatoshi Sato}
\address{Institute for Solid State Physics,
University of Tokyo, 7-22-1 Roppongi, Minato-ku, Tokyo, Japan}

\maketitle
\begin{abstract}
Superconductive properties of the two-leg ladder compounds are studied
theoretically. The antiferromagnetic fluctuations are considered because of
the good nesting  of the Fermi surfaces.
The attractive  interaction which is  most likely due to  the
electron-phonon coupling is also taken into account. Under this
circumstance, it is shown that the superconductivity has two sets of
spin-triplet pairings.
The gaps for both pairings  have no node.
\end{abstract}

]

\narrowtext
A number of unconventional superconductivities are under intensive study in
recent years.
Examples include the high $\Tc$ cuprates, heavy fermions, Sr$_2$RuO$_4$,
organic conductors etc.
The common property of these compounds is the existence of
antiferromagnetic(AF) fluctuations.
Because of this a large number of works are devoted to understand the
unconventional nature from AF fluctuations.
As a result most of them neglect the electron-phonon interactions which
explain most of the superconductors\cite{BCS} before the discovery of
the high $\Tc$ cuprates.
Although, there exist a few attempts to explain the unconventional
nature of superconductivity from  attractive interactions (which most
likely come from electron-phonon coupling) under the influence of AF
fluctuations\cite{FK,CFK,KS,SK}.

Recently  the ladder compounds which may have
some connections with the high $\Tc$ cuprates were found.
The crystal structures are different
but the basic role of Cu and O looks similar.
Hiroi et al.\cite{hiroi} were the first to
synthesize the family of layer compounds Sr$_{n-1}$Cu$_{n}$O$_{2n-1}$, which
have arrays of parallel line defects.
Nearly ideal ladder compounds should result.
The first member ($n=2$ or SrCu$_{2}$O$_{3}$) has two-leg ladders, the
second ($n=3$ or Sr$_{2}$Cu$_{3}$O$_{5}$) has three-leg ladders and so
on.
It was also clearly shown that the material LaCuO$_{2.5}$ has an
insulator-metal transition upon hole doping by substitution of Sr$^{2+}$
for La$^{3+}$ but no sign of superconductivity was observed\cite{hiroi2}.
There have been considerable interests in magnetism
of the ladder compounds \cite{dagotto}.

The ladder material (Sr, Ca)$_{14}$Cu$_{24}$O${_{41-\delta}}$ was
synthesized by MaCarron et al.\cite{ladder1} and by Siegrist et
al.\cite{ladder2}.
In this compound superconductivity was found by Uehara et
al.\cite{super}.
It appears at around $\Tc \sim 10$K but only under high pressure more
than 3GPa.
Due to this limitation essential properties of these compounds have not
yet clarified by experiments.
In particular the nature of superconductivity is not well understood.

In this paper we study  the superconductive properties of
two-leg(two-chain) ladders theoretically.  In order to obtain universal
features we only consider the minimum model which is supposed to give
essential features of superconductivity.  Details which are pertinent to
specific compounds are not considered.  Neither did we try to
estimate the values of physical quantities like $\Tc$.
This is because, for quantitative predictions, one needs to fix various
physical quantities which are not known  theoretically or
experimentally. 
Although, as shown below, $\Tc$ is similar to those of the classical BCS
superconductors.

The existence of unconventional spin-triplet superconductivity in the
ladder systems is shown.
This result seems to be robust and there is a possibility to find
unconventional superconductivity in other ladder materials which include
more than two legs ladders.

\noindent
--{\em Band structure} \\
The energy dispersion of the noninteracting  two-chain ladder
is given

\begin{eqnarray}
\varepsilon(k)=-2t\cos k_x \pm t' ,
\label{ek}
\end{eqnarray}
where $t$ is the transfer integral in the $x$-direction which is along the
chains (usually denoted as $c$-axis direction), and $t'$ is the transfer
integral between the two chains. The two transfer integrals, $t$ and $t'$ have
the same order of magnitude.  The sign in front of $t'$ represents parity with
respect to  exchange of the two chains.

The inter-ladder hoppings are small and estimated to be $t'' \sim 1/20  \ {\rm
to} \ 1/30 \ t'$, and we shall neglect them hereafter.
\begin{figure}
\centerline{\epsfxsize=4.5cm\epsfbox{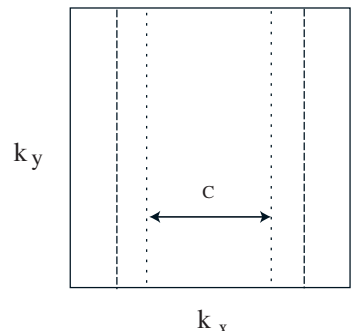}}
\caption{The Fermi surfaces in the first Brillouin zone. The broken
lines are the Fermi surfaces of the  even-parity (with respect to
exchange of the two chains) lower band and the dotted lines are the
Fermi surfaces  of the  odd-parity upper band.}
\label{fig1}
\end{figure}

There are four Fermi surfaces as shown in Fig. \ref{fig1}.
The superconductive pairs with finite momenta are in general have
smaller binding energies compared with those with momentum zero, {\it
i.e.} those consist of electrons in the same band (either
$+$ or $-$ in Eq.(\ref{ek})). Therefore we shall consider only one band
and the two Fermi surfaces for the moment.

\begin{figure}
\centerline{\epsfxsize=4.5cm\epsfbox{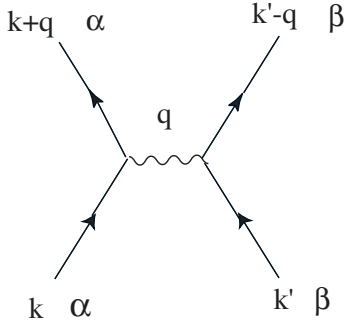}}
\caption{Interaction which does not involve spin. The spins does not
 change by the interaction.}
\label{fig2}
\end{figure}

\noindent
--{\em Attractive interaction} \\
First let us study interactions which are attractive and do not change
spins. See Fig. \ref{fig2}. The general form is
\begin{eqnarray}
H_{{\rm int}}=-
\sum_{k,k'}\sum_{\alpha,\beta}
:f(q) a_{k+q,\al}^\dagger
a_{k,\al}
a_{k'-q,\bt}^\dagger a_{k',\bt}: ,
\end{eqnarray}
where $f(q)>0$ represents an attractive interaction and mainly depends on
$q_x$.
It has a peak at $q_x=0$. 
Here $:\,:$ denotes creation-annihilation normal ordering.
This interaction most likely comes
from electron-phonon coupling. Let us concentrate on the interactions
which are
relevant to
$(k,-k)$ pairing.
Thus choose $k' =-k$ and put $q=k'' - k$
\begin{eqnarray}
H_{ {\rm int}}=
\sum_{k,k''}\sum_{\alpha,\beta}
f(k''-k)
a_{k'', \al}^\dagger a_{-k'', \bt}^\dagger
a_{k,\al}a_{-k,\bt}. \nonumber \\
\label{hint}
\end{eqnarray}
Let us  introduce $V_{s_1s_2s_3s_4}(k,k')$ by
\begin{eqnarray}
&&H_{{\rm int}} \nonumber \\
&&={1 \over 2}
\sum_{k,k'}\sum_{s_1,s_2,s_3,s_4}
V_{s_1s_2s_3s_4}(k,k')
a_{-k,s_1}^\dagger a_{k,s_2}^\dagger a_{k',s_3} a_{-k',s_4},
\label{hint2}
\end{eqnarray}
where $s_i's$ are spin indices.
By the symmetry
\begin{eqnarray}
V_{s_1s_2s_3s_4}(k,k')&=&-V_{s_2s_1s_3s_4}(-k,k')
\nonumber\\
&=& -V_{s_1s_2s_4s_3}(k,-k'),
\label{sym}
\end{eqnarray}
we obtain
\begin{eqnarray}
&&V_{s_1s_2s_3s_4}(k,k')
\nonumber\\
&&={1\over2}\{f(k+k') +
f(-k-k'))\}\delta_{s_1s_3}\delta_{s_2s_4}
\nonumber \\
&&-{1\over2} \{f(k-k') + f(-k+k')\} \delta_{s_1s_4}\delta_{s_2s_3}.
\label{vint}
\end{eqnarray}
\begin{figure}
\centerline{\epsfxsize=4.5cm\epsfbox{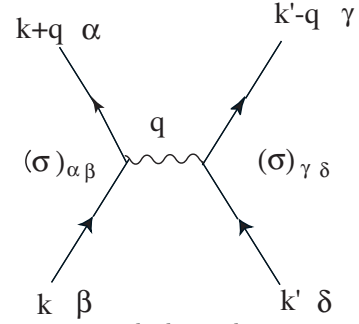}}
\caption{
Interaction which involves spins. The spins can changed by the interaction.
}
\label{fig3}
\end{figure}
\noindent
--{\em Magnetic  interaction} \\
The general form of magnetic interaction is
written (see Fig. \ref{fig3})
\begin{eqnarray}
&&H_{\rm int}^{\rm m}
\nonumber\\
&&=-\sum_{k,k',q}\sum_{\alpha,\beta,\gamma,\delta}
:J_x(q) (\sigma_x)_{\al\bt}\cdot (\sigma_x)_{\gamma\delta}
a_{k+q,\al}^\dagger a_{k,\bt}
a_{k'-q,\gamma}^\dagger a_{k',\delta}:
\nonumber\\
&&-\sum_{k,k',q}\sum_{\alpha,\beta,\gamma,\delta}
:J_y(q) (\sigma_y)_{\al\bt}\cdot (\sigma_y)_{\gamma\delta}
a_{k+q,\al}^\dagger a_{k,\bt}
a_{k'-q,\gamma}^\dagger a_{k',\delta}:
\nonumber\\
&&-\sum_{k,k',q}\sum_{\alpha,\beta,\gamma,\delta}
:J_z(q) (\sigma_z)_{\al\bt}\cdot (\sigma_z)_{\gamma\delta}
a_{k+q,\al}^\dagger a_{k,\bt}
a_{k'-q,\gamma}^\dagger a_{k',\delta}: .
\nonumber\\
\label{hm}
\end{eqnarray}
In general $J_i(q)(>0)$'s ($i=x,y,z$) are different each other,
but for simplicity we treat explicitly only the case
$J_x(q)=J_y(q)=J_z(q)\equiv J(q)$ in the following.
The anisotropy of the magnetic interaction amplifies the tendency toward
the odd-parity superconductivity, and it determines the direction of the
$\vec{d}$ vector ( which is defined by Eq.(\ref{dvec}))
\cite{SK}. If the anisotropy is strong enough, the odd-parity
superconductivity is realized even if $f(q)=0$.

By considering only the terms which contribute ($k, -k$) pairings, we have (by
putting $k=-k'$)
\begin{eqnarray}
&&H_{{\rm int}}^{\rm m}
\nonumber\\
&&= -\sum_{k,q}\sum_{\alpha,\beta,\gamma,\delta}
:J(q)({\vec \sigma})_{\al\bt}\cdot ({\vec \sigma})_{\gamma\delta}
 a_{k+q, \al}^\dagger a_{k,\bt} a_{-k-q,\gamma}^\dagger a_{-k,\delta}:.
\nonumber\\
\end{eqnarray}
Put $q=k''-k$ then
\begin{eqnarray}
&&H_{{\rm int}}^{\rm m} \nonumber \\
&& = -\sum_{k,k''}\sum_{\alpha,\beta,\gamma,\delta}
:J(k''-k)({\vec \sigma})_{\al\bt}\cdot ({\vec \sigma})_{\gamma\delta}
a_{k'',\al}^\dagger  a_{k,\beta}
 a_{-k'',\gamma}^\dagger a_{-k,\delta}: \nonumber \\
&& = \sum_{k,k''}\sum_{\alpha,\beta,\gamma,\delta}
J(k''-k) ({\vec \sigma})_{\al\bt}\cdot ({\vec \sigma})_{\gamma\delta}
a_{k'',\al}^\dagger  a_{-k'',\gamma}^\dagger
a_{k,\beta} a_{-k,\delta}. \nonumber \\
\label{hintm}
\end{eqnarray}
By comparing this equation (\ref{hintm}) and Eq.(\ref{hint2}), and using the
symmetry (\ref{sym}), the magnetic interaction is written
\begin{eqnarray}
&&V_{s_1s_2s_3s_4}^{\rm m}(k,k') \nonumber \\
&&={1\over 2}\{J(k+k')+J(-k-k')\}
({\vec \sigma})_{s_1s_3}\cdot ({\vec \sigma})_{s_2s_4}
\nonumber \\
&&-{1\over 2}\{J(k-k')+J(-k+k')\} (
{\vec \sigma})_{s_1s_4}\cdot ({\vec \sigma})_{s_2s_3}.
\label{vmint}
\end{eqnarray}

\noindent
--{\em Gap equation}\cite{BCS}  \\
We use the weak coupling BCS theory. It is possible that the strong-coupling
corrections modify the  results  quantitatively but not qualitatively because
$\Tc$ is not so high. The gap equation is
\begin{eqnarray}
\Delta_{ss'}(k)
=&-&\sum_{k',s_3,s_4}V_{s'ss_3s_4}(k,k')
\nonumber\\
&\times&{{\Delta_{s_3s_4}(k')} \over{2E_{k'}}}
\tanh \left({\bt E_{k'} \over 2}\right),
\label{gapeqn}
\end{eqnarray}
where the (scalar) excitation energy $E_{k'}$ is given by
\begin{eqnarray}
E_k= \sqrt{\varepsilon (k)^2
+\frac{1}{2}{\rm tr}(\Delta(k)\Delta(k)^{\dagger})},
\end{eqnarray}

For even-parity states (spin-singlet pairing) one can write \cite{leggett}
\begin{eqnarray}
\Delta(k)&=&i\sigma_y\psi(k)
\end{eqnarray}
with $\psi(k)=\psi(-k)$.
Thus the gap equation for  even-parity states is
\begin{eqnarray}
 \psi(k) = -\sum_{k'}
V_{kk'}^{\rm even}
\frac{\psi(k')}{2E_{k'}}
\tanh\left(
{{\bt E_{k'}}\over {2}}
\right).
\label{gapeven1}
\end{eqnarray}
where $V_{kk'}^{\rm even}$ is obtained from  Eq.(\ref{vint})  and
Eq.(\ref{vmint}) as
\begin{eqnarray}
&&V_{kk'}^{\rm even}
\nonumber\\
&& =-{1\over2}\{f(k+k')+f(-k-k') +f(k-k')+f(-k+k')\}
\nonumber
\\
&&
+\frac{3}{2} \{ J(k+k')+J(-k-k')+J(k-k')+J(-k+k') \}.
\nonumber
\\
\label{veven}
\end{eqnarray}

For odd-party states (spin-triplet pairing) one can write \cite{leggett}
\begin{eqnarray}
\Delta(&k&)=i(\vec{d}(k)\cdot {\vec \sigma })\sigma_y \nonumber \\
\label{dvec}
\end{eqnarray}
 with
$\vec{d}(k)=-\vec{d}(-k)$.
The gap equation for odd-parity states is
\begin{eqnarray}
&\vec{d}(k)& = -\sum_{k'}
V_{kk'}^{\rm odd}
\frac{\vec{d}(k')}{2E_{k'}}
\tanh\left({{\bt E_{k'}}\over {2}}\right)
.
\label{gapodd}
\end{eqnarray}
where $V_{kk'}^{\rm odd}$ is obtained from  Eq.(\ref{vint})  and
Eq.(\ref{vmint}) as
\begin{eqnarray}
&&V_{kk'}^{\rm odd}
\nonumber\\
&&={1\over2} \{f(k+k')+f(-k-k')- f(k-k')-f(-k+k')\}
\nonumber \\
&&+\frac{1}{2}\{ J(k+k') +J(-k-k')-J(k-k')- J(-k+k')\}.
\nonumber\\
\label{vodd}
\end{eqnarray}

\noindent
--{\em BCS approximation}\cite{BCS}\\
Following the BCS treatment of superconductivity we make further
approximations. Namely the pairing interactions
are  constant  within the cutoff and zero outside the cutoff.
The function $f(q)$ has a peak at
$q_x=0$ but it is approximated by a constant $V >0$. Therefore this
interaction corresponds to the phonon-mediated interaction in the
classical BCS theory.  In addition to
this interaction the ladder compounds have   interactions due to AF
fluctuations which are induced by nesting of the Fermi surfaces in a similar
fashion to some of the organic conductors or the high $\Tc$ cuprates
in the low doping region.  Since the
Fermi surfaces are straight
and parallel, the nesting is perfect. Therefore the nesting vector is $Q=(c,
k_y,k_z)$
where {c} is the distance between the two Fermi surfaces of the same band. See
Fig.1. (Nesting is also perfect between the Fermi surfaces of different
bands. But these
are not relevant to the superconducting pairing of $(k, -k)$.) Therefore
there exist AF
fluctuations with nesting vector
$Q$. Now we take the magnetic interaction $J(q)$ to be constant $J(Q)>0$
when $q$ is is close to $Q$ and connects two points in different Fermi
surfaces (of the
same band) within the cutoff. Then Eq.(\ref{veven}) gives
\begin{eqnarray}
V^{\rm even} = -V+3J(Q)
\end{eqnarray}
for even-parity states, and  Eq.(\ref{vodd}) gives

\begin{eqnarray}
V^{\rm odd} = -V+J(Q)
\label{oddint}
\end{eqnarray}
for odd-parity states.
Therefore we always have $-V^{\rm odd} > -V^{\rm even}$ as long as  magnetic
interactions are present. It leads to odd-parity (spin-triplet)
pair condensation with
\begin{eqnarray}
T_{\rm c} = 1.13\omega_{\rm D} e^{1/\{N(k_{\rm F}) V^{\rm odd}  \}}.
\end{eqnarray}
where $\omega_{\rm D}$ is a cutoff.

When the anisotropy of AF fluctuations is taken into account,
$V^{\rm odd}$ is replaced by
\begin{eqnarray}
V^{\rm odd}=
\left\{
\begin{array}{ll}
-V-J_x(Q)+J_y(Q)+J_z(Q)& \mbox{for $d_x\neq 0$}\\
-V-J_y(Q)+J_z(Q)+J_x(Q)& \mbox{for $d_y\neq 0$}\\
-V-J_z(Q)+J_x(Q)+J_y(Q)& \mbox{for $d_z\neq 0$}
\end{array}
\right. .
\label{oddint2}
\end{eqnarray}
It can be seen from Eq.(\ref{oddint2}) that if AF fluctuations
perpendicular to the $\vec{d}$ vector are strong enough, there is no
superconductive condensation.
It also can be seen that AF fluctuations parallel to the $\vec{d}$
vector promote the odd-parity superconductivity.
It is rather straightforward to
show that there is no node in the gaps from the gap equation
(\ref{gapeqn}).

\noindent
--{\em Discussions} \\
To conclude, we study the unconventional superconductive nature of the
two-leg ladder compounds. The attractive interactions
which are most likely due to the electron-phonon coupling are considered.
AF fluctuations from the  good nestings  of the Fermi surfaces  are also
included.

Our main conclusion is that the two-leg ladder compounds have two sets of
spin-triplet superconductive states without nodes in the gap
corresponding the two sets of the Fermi surfaces.
These superconductivities are  suppressed by AF fluctuations perpendicular
to the $\vec{d}$ vector and
vanish if the AF fluctuations are strong enough.
Our results also apply to the ladder compound with more than two legs.

It is possible to have two $\Tc$'s corresponding to the two bands,
respectively. However it is likely to have a single $\Tc$ due to some proximity
effects induced by the interactions which are not included here.

\begin {center}
{\large{\bf Acknowledgments}}

\end {center}
It is a pleasure to thank J. Akimitsu and N. Mori for  useful discussions.


\begin{references}


\bibitem{BCS} J. Bardeen, L.N. Cooper,and J.R. Schrieffer, Phys. Rev. {\bf
106}, 162;
{\bf 108}, 1175 (1957).
\bibitem{FK}
J. Friedel and M. Kohmoto, (preprint) cond-mat/9901065.
\bibitem{CFK}
I.Chang, J. Friedel and M. Kohmoto, (preprint) cond-mat/9908214.
\bibitem{KS}
M. Kohmoto and M. Sato, (preprint) cond-mat/0001331.
\bibitem{SK}
M. Sato and M. Kohmoto, (preprint)  cond-mat/0003046.
\bibitem{hiroi} Z. Hiroi, M. Azuma, M. Takano, and Y. Bando, J. Solid
State Chem. {\bf 95}, 230 (1991).

\bibitem{hiroi2} Z. Hiroi and M. Takano,
Nature {\bf 377}, 41 (1995).

\bibitem{dagotto} For a review, see {\it e.g.} E. Dagotto and T.M. Rice,
Science {\bf 271}, 618 (1996).

\bibitem{ladder1} E.M. McCarron, M.A. Surbramanian, J.C. Calabrese, and R.L.
Harlow, Mater. Res. Bull. {\bf 23}, 1355 (1988).

\bibitem{ladder2} T. Siegrist, L.F. Schneemeyer, S.A. Sunshine, J.V.
Waszczah and R.S. Roth, Mater. Res. Bull. {\bf 23}, 1429 (1988).

\bibitem{super} M. Uehara, T. Nagata, J. Akimitsu, H. Takahashi, N. Mori, and
K. Kinoshita, J. Phys. Soc. Jpn. {\bf 65}, 2764 (1996).

\bibitem{leggett}
See {\it e.g.} A. Leggett, Rev. Mod. Phys. {\bf 47}, 331 (1975);
M. Siegrist and K. Ueda, Rev. Mod. Phys. {\bf 63}, 239 (1991).


\end{references}
\end{document}